\documentclass[9pt,twocolumn,twoside]{opticajnl}
\journal{opticajournal}
\setboolean{shortarticle}{false}

\title{Fully stabilized Er fiber comb at 1 GHz by harmonic modelocking}

\author[]{Kevin F. Lee}
\author[]{Jacob Lampen}
\author[]{Peng Li}
\author[]{Jie Jiang}
\author[]{Martin E. Fermann}
\affil[]{IMRA America, Inc. 1044 Woodridge Ave, Ann Arbor, MI 48105, USA}

\begin{abstract}
Modelocked frequency comb lasers have always operated with a single pulse circulating in the laser cavity.  This meant that each laser technology had an associated limit on pulse repetition rate.  Achieving higher rates required different technology, for example exchanging optical fiber for solid state gain media.  However, this is a perceived, rather than a fundamental limit.  We demonstrate a new fiber laser design with multiple pulses circulating in the fiber gain cavity with the same high precision as a conventional fiber frequency comb.  This has the immediate benefit of bringing mature fiber technology to higher repetition rate frequency combs.  More generally, it adds great design freedom to laser engineering, where the laser can be separated into an optical cavity and a gain medium that are combined using standard frequency comb techniques.  Fundamental frequency comb performance improvements may even be possible from the filtering intrinsic to our design, or by incorporating high stability cavities directly into the laser itself.
\end{abstract}

\setboolean{displaycopyright}{false} 

\begin{document}

\maketitle

\section{\label{secIntroduction}Introduction}

\begin{figure*}
\includegraphics[width=183mm]{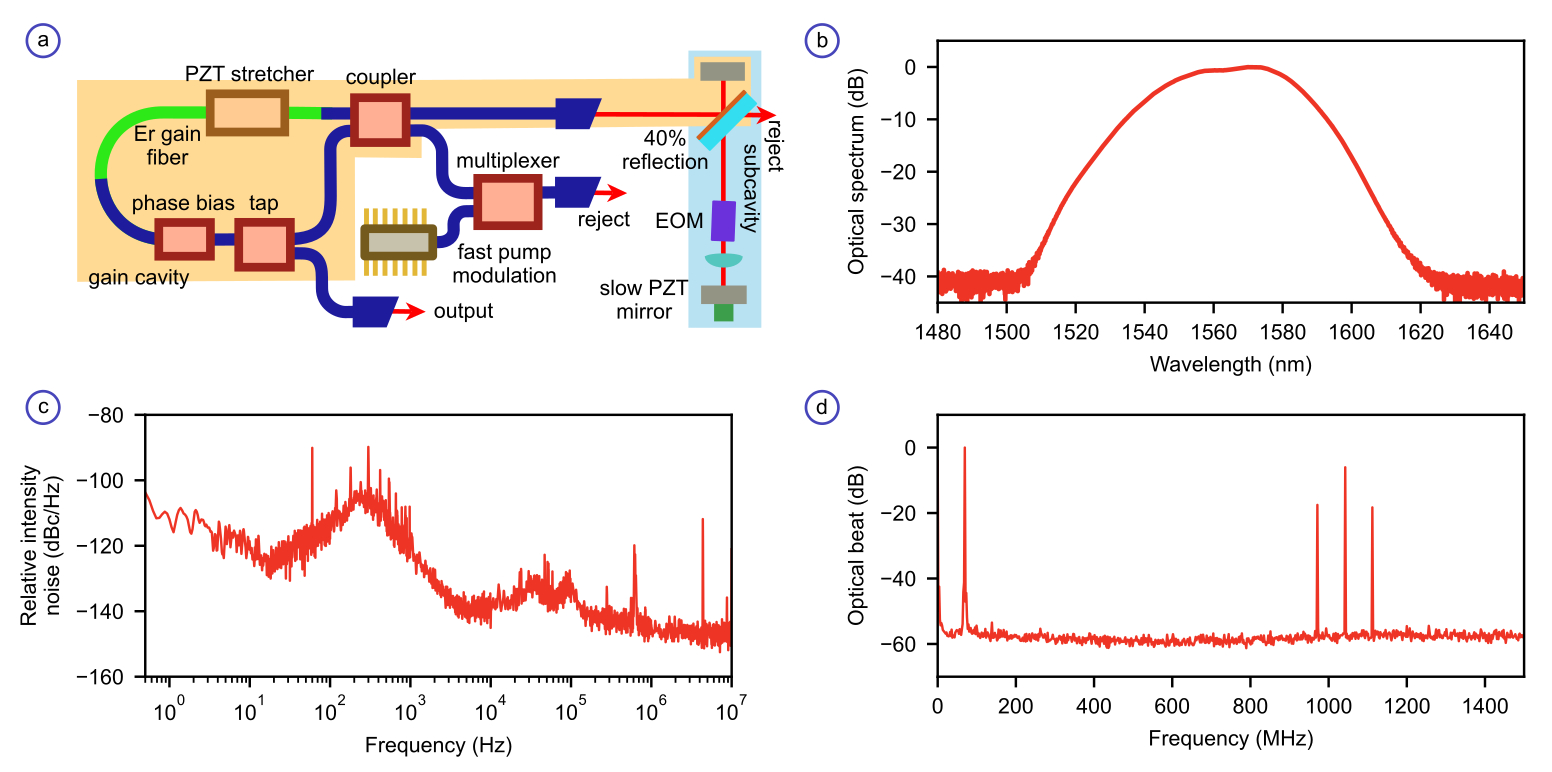}
\caption{\label{figLaser}
GHz laser showing basic characteristics quite similar to an ordinary fiber comb.  (a) Drawing of harmonically modelocked GHz fiber laser design using a short freespace subcavity (right, blue shade) to increase repetition rate of a conventional NALM fiber oscillator (left, orange shade).  (b) Spectrum of laser output at 1 GHz repetition rate supporting femtosecond pulses.  (c) Relative intensity noise of 1 GHz comb, with some noise around 300 Hz.  (d) In-loop optical beating of GHz comb with a CW laser (30 kHz RBW).  The lack of extraneous peaks indicates clean harmonic modelocking.
}\end{figure*}

Femtosecond frequency combs are a core technology of modern photonics.  Combs provide short pulses for nonlinear effects, precision frequency transfer and timing for metrology, bandwidth for multiplexing, and high precision timing for synchronization and high data throughput \cite{changNPh2022}.  In time and frequency metrology, one of the most demanding of applications, fiber lasers are the platform of choice \cite{fortierCP2019}, with reliable, mature technology, and precise transfer and timing performance.  

Until now, fiber frequency combs operated with one pulse circulating in the laser.  Given fiber lengths of about a meter in an erbium fiber oscillator, fiber comb repetition rates are usually in the neighbourhood of 200 MHz.  Different applications may need higher rates, such as few or tens of GHz for comb resolution in spectroscopy or astronomy \cite{wilkenN2012}, or photonic communications \cite{corcoranNComm2020}.

To produce GHz combs in laboratory applications like astronomy, fiber combs are scaled up externally by spectral filtering and pulse multiplication in high reflectivity optical cavities, at the cost of lost laser power in the filtered comb lines, and an extra cavity lock.  Fiber combs have been pushed to GHz rates with short, highly doped Yb fibers \cite{hartlCLEO2009, jiangOE2014}, but these specialized, non-polarization maintaining systems do not share the technological maturity of erbium combs, and are not scalable to higher rates.  Otherwise, different laser technologies can be used \cite{fortierCP2019}, such as solid state around 1 GHz \cite{leskoOSAC2020, mullerOE2023}, electrooptic combs \cite{zhangN2025}, or microresonators at tens of GHz \cite{gaetaNPh2019}, but this forgoes the benefits of fiber laser technology such as ultralow noise, power scalability, large spectral range, and low cost (at least until chip device costs decrease).

We instead aim to have many pulses in the laser at once to scale up the repetition rate while using established fiber technology.  Assuming enough pumping to have the same pulse energy at any repetition rate, the nonlinear effects on each pulse match those for a single circulating pulse, so multiple pulses can exist independently within the laser.  Noisy multipulsing is common when starting a modelocked laser.  Clean operation with a repetition rate at a multiple of the free spectral range is called harmonic modelocking.

Harmonic modelocking has been researched for over 50 years \cite{beckerJQE1972} and seen in ps fiber lasers over 30 years ago \cite{kafkaOL1989} at GHz pulse rates \cite{takadaEL1990}.  Harmonically modelocked lasers are prone to including unwanted lasing modes, called supermode noise.  Supermodes can be suppressed by adding spectral filtering with a Fabry Perot etalon \cite{harveyOL1993}, which is actually a GHz subcavity within the laser cavity.  Traces of supermodes can remain though, even with Pound Drever Hall (PDH) stabilization \cite{geePTL2004}, manifesting as traces of the free spectral range spacing in radio frequency (RF) spectra \cite{hasegawaOE2024, korobkoOE2022}.  More recently, microresonators \cite{rowleyN2022}, small fiber loops \cite{caoLPOR2023} and fiber knots \cite{dingAdvPh2020} have been used as subcavities for high repetition rates.  Comb properties are starting to be measured \cite{hasegawaOE2024, caoLPOR2023, cutronaAPL2023}, and fast actuation for full comb stabilization is being pursued \cite{cutronaAPL2023}.

For clean frequency combs, pulses need to be coherent with each other, implying strong coupling between adjacent pulses, which does not occur with harmonic modelocking in a standard oscillator.  In our new design, we separate the functions of a laser into two cavities as illustrated in Fig. \ref{figLaser}a.  A smaller free space subcavity (blue shade) governs the repetition rate, with a single pulse oscillating inside, ensuring pulse to pulse coherence.  Gain and modelocking of the subcavity is outsourced to a much longer nonlinear amplifying loop mirror (NALM) fiber gain cavity (orange shade) \cite{fermannOL1990, kuseOE2016}.  Since the 1 GHz subcavity is 15 cm long, and the gain cavity length is not restricted by the repetition rate, actuators can be easily added for comb stabilization.

A beamsplitter couples the subcavity to the tail end of the gain cavity.  If the lower half of the subcavity is blocked, the system can run as a normal oscillator.  In operation, the subcavity pulse must meet a returning amplified pulse at the beamsplitter to interferometrically cancel the beam leaving the laser.  The round trip time of the gain cavity must be an integer multiple of the subcavity round trip time so the subcavity pulse will effectively have gain when going up through the beamsplitter by joining an amplified pulse from the gain cavity.    

Interference at the beamsplitter is a highly selective filter, presenting an opportunity for improving performance.  The beamsplitter is effectively a large loss for amplified light that does not match light in the subcavity in wavelength, phase, or timing.  In contrast, an inline method like a Fabry-Perot etalon \cite{harveyOL1993} does not have a rejection port, so noise remains in the laser.  Filtering and harmonic modelocking reduces fundamental noise limits, particularly timing jitter, which we discuss in the Supplement.

The subcavity could also be viewed as providing pulsed self-injection locking of the fiber laser.  Conventional self-injection locking narrows the linewidth of continuous wave (CW) lasers by filtering and reflecting some output back as a time-delayed seed \cite{dahmaniOL1987}.  It is used in common pump diodes, and high precision commercial narrow linewidth sources \cite{malekiLFW2014}.  Our system also couples the laser field to a past version of itself.  Diode lasers easily adapt to weak seeding, while our system uses very strong feedback that can completely change the pulse in the laser.

Our main design operates at 10 × 104 MHz = 1.04 GHz, which is easily powered by two 1 W diodes (details in the Supplement).  This laser has run at repetition rates from the fundamental to the 12th harmonic simply by changing the subcavity length, giving it over an order of magnitude of range.  We also made design variants with different subcavities.  In one, we begin fiberizing the subcavity to improve reliability and size for practical applications.   In another we have operated a one-piece, empty subcavity to act as a passively stable internal reference to eventually bring some of the stability of high finesse reference cavities \cite{oelkerNPh2019} directly into the oscillator.

\section{\label{secComb}Comb stabilization}

\begin{figure*}
\includegraphics[width=183mm]{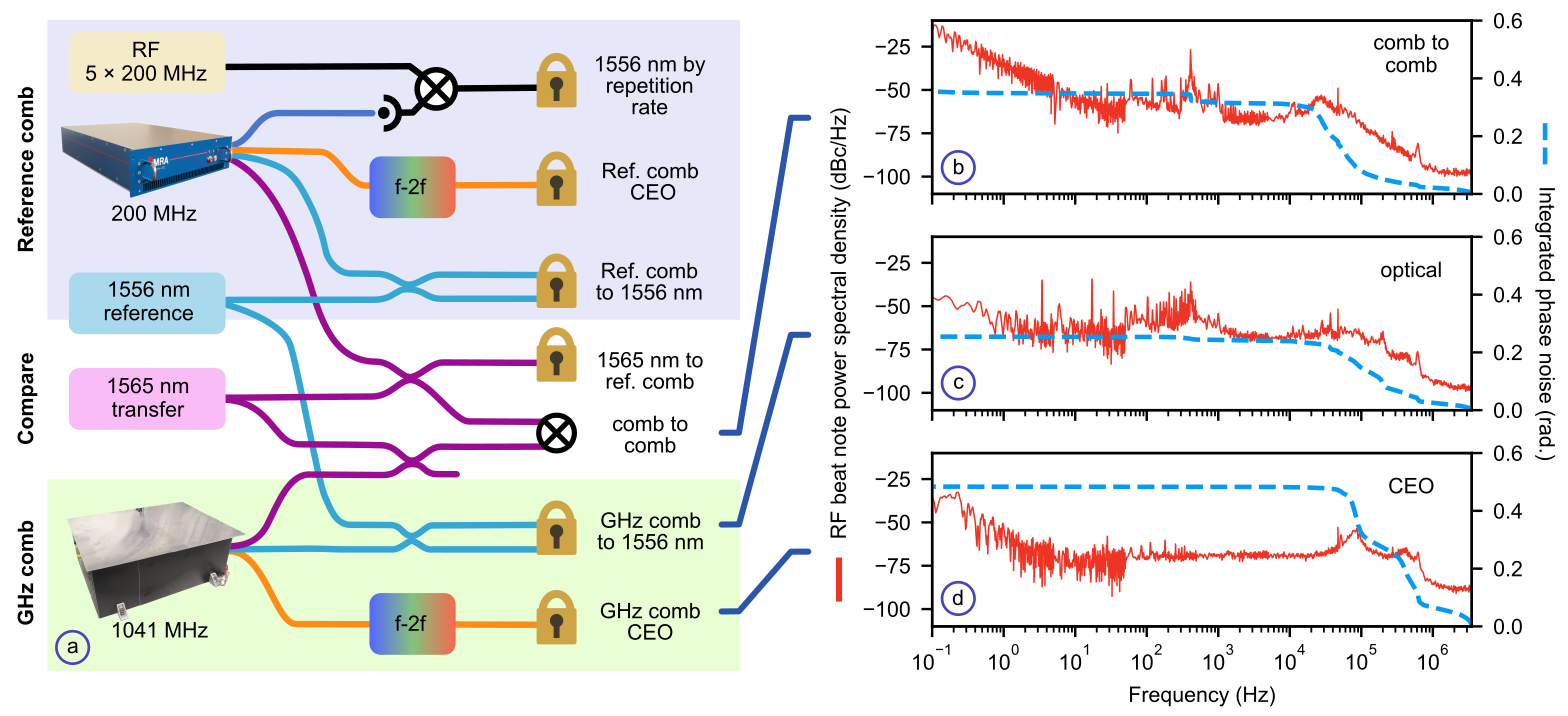}
\caption{\label{figBeats}
(a) locking and interference arrangement for harmonically modelocked GHz comb including a standard fiber comb for comparison.  Both combs have stabilized CEO and are locked to a 1556 nm optical reference which is stabilized against the reference comb repetition rate.  The distance between comb lines from each comb is measured by mixing the beat signals of each comb with a 1565 nm laser which is locked to a reference comb line.  Right: RF spectrum and integrated phase noise for the: (b) out-of-loop comb to comb beat note (0.36 rad); (c) the locked in-loop optical beat note (0.23 rad); and (d) CEO beat note (0.48 rad).  The comb to comb beat has a linewidth below 0.1 Hz.  These values are all comparable to standard fiber comb performance, and verify stable GHz frequency comb generation from our harmonically modelocked laser.
}\end{figure*}

The initial reaction to a multicavity laser might be to assume instability as the cavities vibrate and drift in and out of phase.  This is somewhat true, but locking cavities to achieve new capabilities is common practice with combs \cite{leeJPB2016}.  To stabilize, the subcavity is tuned by a fast electrooptic modulator (EOM) and a slow mirror on a piezoelectric transducer (PZT).  The gain cavity is tuned by fast pump current modulation and a slow PZT fiber stretcher.  For non-comb applications, maintaining modelocking is not difficult.  The system in Fig. \ref{figLaser}a, with a 40\% reflectivity coupler, is almost passively stable.  Since the spectral bandwidth depends strongly on the relative cavity length, the output power in a spectral window on the side of the spectrum is a good diagnostic of the relative cavity length.  Stabilizing this by feedback at $\sim$10 Hz bandwidth to the fiber stretcher is a simple way to keep the system running, though with drifting comb properties. 

For comb stability, we measure the two comb values: the carrier envelope offset (CEO) frequency by f-2f interferometry; and the repetition rate, indirectly through an optical beat with a narrow linewidth CW reference laser, just as for a conventional comb.  In standard fiber combs, the repetition rate is controlled by fiber length, and the CEO is controlled by adjusting the optical gain or loss \cite{kuseOE2016}.  

In our design, there are two cavities with their own lengths and losses, which suggests four parameters for controlling two comb properties.  The repetition rate is conceptually simple, both cavity lengths should match the desired repetition rate.  The CEO is less obvious, but importantly, we find that the f-2f frequency is very dependent on the relative matching between cavities.

Knowing this, we treat the CEO as a measurement of cavity matching.  We assign the CEO to the pump current and fiber stretcher.  This would be unusual in a conventional comb, but here it controls the CEO by the phase and spacing of pulses returning to the beamsplitter.  We assign the optical beat to the subcavity length, using the EOM and the mirror PZT, while the gain cavity follows along through the CEO lock.  This removes the need for a PDH or other kind of independent lock between the two cavities which we have seen causes crosstalk and breaks modelock as the two target conditions drift apart.

We coherently locked the f-2f and optical beat notes, producing a fully stabilized frequency comb.  Even with an additional cavity, our actuator count is the same as a conventional comb, with a slow and a fast actuator each for the CEO and repetition rate.

\section{\label{secLaser}Laser characterization}

Besides the higher rate, the output of our harmonically modelocked laser is similar to a conventional fiber laser.  Fig. \ref{figLaser}b shows a spectrum when running at 1 GHz, with the spectral width and centre like a traditional erbium NALM oscillator.  Fig. \ref{figLaser}c shows the relative intensity noise (RIN) of the fully stabilized GHz comb.  The noise profile is fairly good, although there is a noticeable peak around 300 Hz.  For low amplitude noise applications like radio frequency (RF) generation, we expect much of the noise can be removed by a standard noise eater after the oscillator.

The RF spectrum of the optical beat note is shown in Fig. \ref{figLaser}d, at 30 kHz resolution bandwidth (RBW), with the comb locked.  There is a repetition rate peak at 1041 MHz, the locked optical beat at 70 MHz, and two replicas around 1 GHz.  The beat note is a sensitive heterodyne measurement of laser noise, and this RF spectrum is clean.  If the laser is not harmonically modelocking well, say by poor length matching, obvious additional optical beats appear, and there may be a narrow peak in the optical spectrum, or strongly modulated output power.

Fig. \ref{figBeats}c shows the locked optical beat note at higher resolution and with the corresponding integrated phase noise from 3.5 MHz of 0.23 rad, comparable to a good fiber comb, though the best NALM oscillators with better actuators can be well below 0.1 rad \cite{kuseOE2016}.

Fig. \ref{figBeats}d shows the f-2f beat note and corresponding integrated phase noise, which reaches 0.48 rad.  This is a reasonably good lock quality, comparable to a conventional fiber comb \cite{sinclairRSI2015}.  The CEO lock quality may be worsened by our f-2f interferometer which is amplified to almost 1 W to reach 1 nJ pulse energies for supercontinuum generation (see Supplement) and was not fully optimized.  It had a low signal to noise ratio (SNR) near 35 dB at 100 kHz RBW.  The CEO lock may also be affected by crosstalk, as the optical beat actuators also affect CEO.

\section{\label{secLoop}Out of loop comparison}

\begin{figure}
\includegraphics[width=89mm]{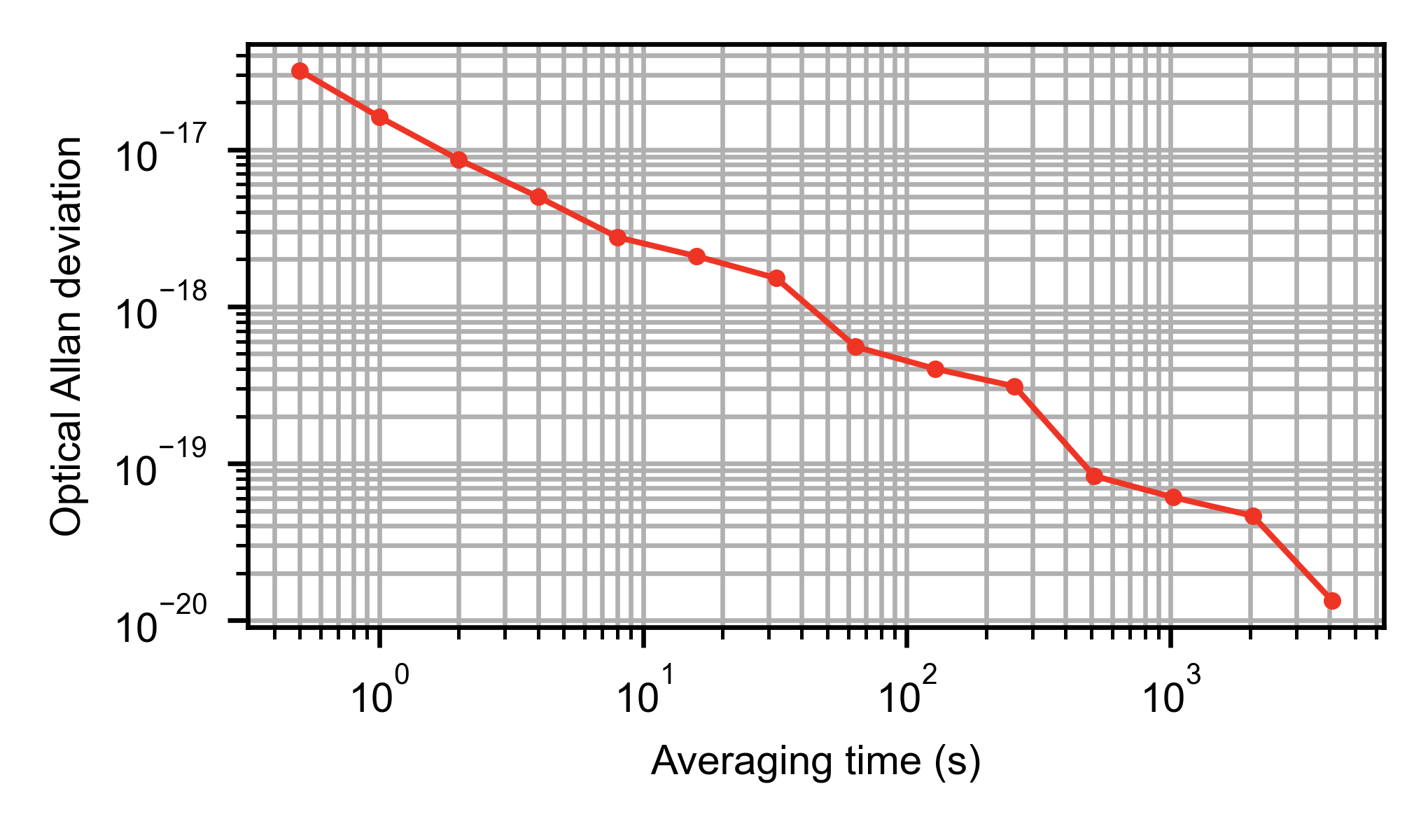}
\caption{\label{figAllanDevs}
Allan deviation of frequency difference between a comb line from the GHz comb and a reference frequency comb.  The deviation decreases steadily, indicating good long term stability.
}\end{figure}

We verify the quality of the frequency comb by comparing against a conventional 200 MHz  reference fiber comb for an out of loop measurement.  As in Fig. \ref{figBeats}a, both combs are stabilized to their own CEO frequencies, and to their optical beats with a narrow linewidth 1556 nm CW laser.  The reference comb repetition rate (5th harmonic at 1 GHz) is stabilized against a Rb standard to reduce drift of both the 1556 nm laser and the reference comb.

Another CW laser at 1565 nm compares comb lines from each comb.  Optical beats with the two combs are electronically mixed, providing the frequency difference between a line from each comb at 1565 nm.  Noise is slightly reduced by locking the 1565 nm laser to the reference comb.  A high resolution RF spectrum of the mixed beat note is shown in Fig. \ref{figBeats}b.  The linewidth is below 0.1 Hz, and the integrated phase noise from 3.5 MHz is 0.36 rad, similar to the phase noise of the locked beat notes of the GHz oscillator, demonstrating strong coherence between the two combs.

For long term performance, we use a phase counter to record the out of loop difference frequency for a few hours.  The counter tracks the phase, noting it every millisecond, and reporting an averaged frequency every 0.5~s.  The frequency evolution is plotted in the Supplement.  The stability of the measured frequency is represented in Fig. \ref{figAllanDevs} as an optical Allan deviation.  This is a measure of the variations in frequency over different time periods, scaled to the optical frequency.  The stability is comparable to conventional fiber combs, steadily averaging down from $10^{-17}$ at 1~s, to $10^{-20}$ in an hour, indicating consistent output with time.

At these high stability levels, the results may be measurement limited.  Environmental changes such as temperature add phase shifts between beams in noncommon fibers, causing drifts in the mixed beat frequency that are unrelated to comb stability.  The few meters of noncommon fibers here were packaged into a lightly insulated box, reducing drifts below 1 Hz.

We find that our harmonically modelocked GHz laser is a high quality frequency comb.  We see clear differences between clean and noisy operation states.  We see no indication of any problems such as frequency noise related to the free spectral range; it behaves very much like a standard frequency comb.

\begin{figure*}
\includegraphics[width=183mm]{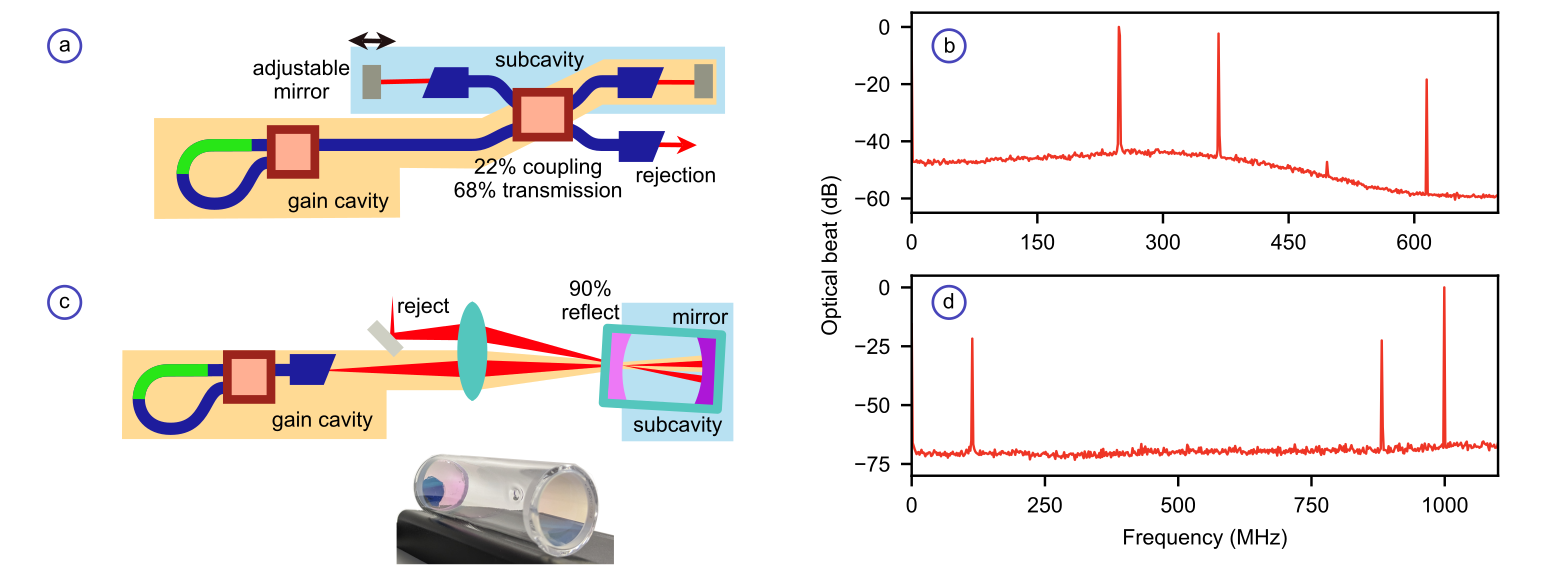}
\caption{\label{figVariant}
Variations of our harmonic modelocking design with different subcavity types.  (a) Fiberizing the subcavity (top, blue shade) to approach an all fiberized GHz comb, but the mirrors are currently still freespace.  The gain cavity is drawn simplified and shaded orange.  (b) The RF spectrum of the optical beating between the fiberized laser and a CW laser.  (c) A double pass V geometry where the subcavity (right, blue shade) is a solid unit, and has no internal optics in order to reduce temperature dependence, as seen in the photograph of the actual cavity.  (d) The RF spectrum of the optical beating of the V cavity laser and a CW laser.  Both lasers show no visible noise peaks in the optical beat (the 496 MHz spike in (b) is an electronic harmonic of the 248 MHz peak).  Both spectra were taken with 100 kHz RBW.
}\end{figure*}

\section{\label{secVariant}Design variants}
While our harmonically modelocked laser runs well, the laser used above is still a laboratory device, with the subcavity on a breadboard.  Sharp taps to the table can break modelock, making it less robust than a fully fiberized oscillator.  Having made a proof of concept laser showing strong comb performance, we now discuss ways we have started improving system stability and capability.

One approach is to fiberize the subcavity.  We previously went from fiber to free space before the beamsplitter.  We have since extended the fiber portion by replacing the beamsplitter with a fiber coupler while still having a free space mirror for length adjustment, as in Fig \ref{figVariant}a.  The fiber coupler nominally transmits 68\% within the same upper or lower branch, and couples 22\% across branches, with a 10\% insertion loss.

We have verified that this design can cleanly harmonically modelock.  We have run it with RF stabilized repetition rate for several hours.  We also measured interference with a CW laser, shown in Fig 4b at 100 kHz RBW, with the repetition rate peak at 615 MHz, and the beat and its replica around 300 MHz.  The small peak at 496 MHz is not laser noise, but the electronic second harmonic of the 248 MHz beat note. 

We plan to fiberize the entire subcavity, mounting all the actuators into a fiberized module as in a standard fiber comb.  A fully fiberized harmonically modelocked comb has the engineering advantages of avoiding optical alignment instabilities, and being packaged into a small box that can be easily stabilized against temperature drifts and vibrations.  

Our other approach aims for a highly stable subcavity that serves as an internal reference.  In metrology, locking a CW laser to an ultrastable high finesse cavity can provide a very high quality optical frequency reference \cite{oelkerNPh2019}.  As a subcavity, the finesse would be lower, as starting and maintaining modelocking is more difficult with lower coupling between cavities.  This will limit the spectral filtering of each resonance, although it is not obvious how the multiple lines over the bandwidth, and laser mode competition affect the laser linewidth.  A higher finesse should also improve the timing jitter, as discussed in the Supplement.

An ultrastable cavity should not have internal optics, as they add temperature dependence.  To move the beamsplitter’s substrate out of the cavity, we use a V shaped subcavity, where a curved beamsplitter is one of the subcavity end mirrors as in Fig. \ref{figVariant}c.  Similar cavities are used in semiconductor injection locking line narrowing \cite{dahmaniOL1987, burkartOL2013}.  Compared to the linear subcavity in Fig. \ref{figLaser}a, we fold the cavity around the beamsplitter coating side, and trap the beamsplitter reflection rather than the transmission so the beamsplitter substrate is outside of the subcavity.  A small reflection angle lets one curved mirror replace the two end mirrors.  The two reflectors were glued to a glass tube to form a solid subcavity.  A clean optical beat with a CW laser is shown in Fig. \ref{figVariant}d, with a repetition rate peak at 1000 MHz showing successful harmonic modelocking with a completely empty subcavity.

The V geometry can also be useful as the subcavity shrinks towards 1 cm size multi-GHz rates.  We imagine using an EOM crystal itself as a solid V subcavity, combining simple construction with high speed actuation.

\section{\label{secConclusions}Conclusions}

We have presented a way to implement harmonic modelocking for generating high quality, fully stabilized frequency combs with fiber lasers at GHz repetition rates.  Though studied for decades, harmonic modelocking was generally seen as adding noise or locking requirements as compared to a conventional laser.  Compared to a standard frequency comb, our GHz laser achieves high precision comb output with the same number of locks, while making repetition rate a new selectable parameter with more than an order of magnitude of range.

Our new GHz fiber combs offer advantages in applications where the repetition rate directly affects performance, like comb-resolved and dual comb spectroscopy.  A dual comb made with polarization duplexing in a shared subcavity with birefringence could provide an ultrastable repetition rate difference.  Our combs are also expected to be a useful tool for photonics and telecommunications as either a source, or a more direct bridge between optical and electronic frequencies, where the potential timing jitter stability improvements would directly improve performance.

Given the new ease of achieving harmonic modelocking, it will be interesting to see this applied to other photonic technologies.  An immediate application would be to fully comb stabilize a microresonator and fiber gain loop \cite{rowleyN2022, cutronaAPL2023}, especially as microcomb actuators continue to improve \cite{liuCLEO2025}.  Harmonic modelocking of a microresonator would remove the CW background, vastly improve pumping efficiency from the conventional 1\% range \cite{liO2022, zhangnpJNanoPh2024}, and add noise filtering.

Novel applications should arise from adapting what are currently external optical cavity techniques.  Our method provides a new way to interact with short external cavities such as spectroscopic enhancement cavities \cite{wangLSA2024}, optical parametric oscillators \cite{allanOCon2024}, and even biosensing cavities like in cellular lasing \cite{gatherNPh2011}, by bringing them right into the oscillator without a power penalty from mismatched repetition rates.

\begin{backmatter}
\bmsection{Acknowledgment} We thank Manfred H. Ober for stimulating discussions.
\end{backmatter}

\section*{}


\clearpage

\setcounter{equation}{0}
\setcounter{figure}{0}
\setcounter{table}{0}
\setcounter{section}{0}
\renewcommand{\theequation}{S\arabic{equation}}
\renewcommand{\thefigure}{S\arabic{figure}}
\renewcommand{\thesection}{S\arabic{section}}
\renewcommand{\bibnumfmt}[1]{[S#1]}
\renewcommand{\citenumfont}[1]{S#1}

\vspace*{5mm}
\begin{center}
{\Huge\bfseries\textsf{Supplement} }
\end{center}
\vspace*{5mm}

We describe a theoretical model of the harmonically modelocked GHz oscillator.  We numerically estimate the reduction of the fundamental timing jitter with changes to the beamsplitter reflectivity.  We also provide some additional experimental details and a plot of the out of loop frequency offset over three hours.

\section{\label{secModel}Model of main GHz oscillator}

In the following we explore the pulse formation process in a harmonically modelocked fiber laser comprising a nonlinear amplifying loop mirror and a subcavity. We present a simple model that describes the salient features of this system and allows predictions for optimized cavity parameters.

\begin{figure}[b] \centering
\includegraphics[width=89mm]{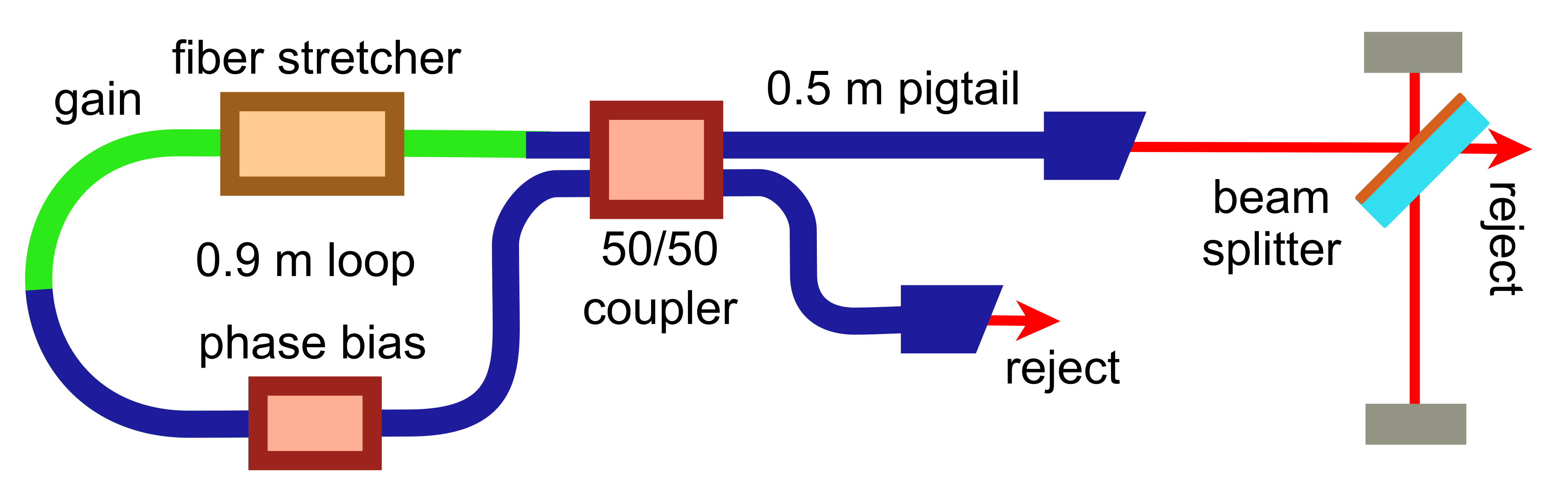}
\caption{\label{figSMtheorylaser}
Simplified version of the harmonically modelocked design used in the simulations.
}\end{figure}

We start with pulse propagation in the nonlinear amplifying loop mirror as shown in Fig. \ref{figSMtheorylaser}.  The NALM has several components, which are all concatenated in the simulation. We start with a pulse with field $A$ entering the NALM pigtail from the right, which is split by the 50/50 coupler into clockwise ($CW$) and counter clockwise ($CCW$) components according to
\[A_{CW} =\frac{i}{\sqrt{2}} A_{in} \]
\[A_{CCW} = \frac{1}{\sqrt{2}} A_{in} \]

After propagation through the loop the output field transmitted through the coupler is obtained from
\[A_{out} = \frac{1}{\sqrt{2}} A_{CW} + \frac{i}{\sqrt{2}} A_{CCW} e^{i\alpha} , \]
where $\alpha \approx \pi/4$ is the non-reciprocal phase bias introduced between the two counter-propagating fields, whereas the output rejected at the coupler is obtained from
\[A_{rej} = \frac{i}{\sqrt{2}} A_{CW} + \frac{1}{\sqrt{2}} A_{CCW} e^{i\alpha} , \]
Propagation of $A_{CC}$ and $A_{CCW}$ can be described by the standard nonlinear Schrödinger equation,
\[ \frac{\partial A}{\partial z} = (g-l) + i \left( \frac{\beta_2}{2} - i \frac{g}{\Omega_g}\right) \frac{\partial^2 A}{\partial t^2} + i \gamma |A|^2 A  \]
where $g, l$ are the distributed gain and loss coefficient, $\Omega_g$ is the gain bandwidth assumed to be parabolic, $\beta_2$ is the second order dispersion parameter and $\gamma$ is the nonlinearity parameter of the fiber. For simplicity we neglect third-order dispersion as well as higher-order nonlinear terms.  To approximately match the experimental configuration, the pigtail length is assumed to be 0.5 m and the loop length 0.9 m, producing a fundamental repetition rate of around 100 MHz. The pigtail and loop $\gamma$ parameters are taken to be 0.01 (mW)$^{-1}$ and 0.013 (mW)$^{-1}$ respectively.  The pigtail and loop $\beta_2$ parameters are taken to be $20000$ fs$^2$/m and  $-23000$ fs$^2$/m respectively; the system is assumed to operate with slightly negative overall dispersion.

The gain coefficient is assumed to saturate according to 
\[ g = g_0 / (1 + E / E_{sat}) \]
where $E$ is the pulse energy, $E_{sat}$ the saturation energy, and $g_0$ the unsaturated gain coefficient.  To account for the asymmetric gain distribution in the loop, we lump all the gain into a single element right after the input is split by the coupler.  Also, a single loss element is inserted to account for the system loss. Quantum noise can further be included according to $P_{ASE} = (G-1) \theta h \nu \Delta\nu$, where $G$ is the overall intensity gain, $\theta \approx 2$ is the excess noise factor, $h$ is Planck’s constant, $\nu$ is the carrier frequency and $\Delta\nu$ is the optical bandwidth. Here the quantum noise introduced is a factor of two higher than spontaneous emission noise due to quantum noise addition through the loss elements inside the laser \cite{SMpaschottaAPB2004}. For a subcavity perfectly matched to the main NALM, the subcavity acts like a perfect mirror, ideally adding no additional quantum noise.

When seeded with a single pulse field $A_{out}$, the intracavity field evolves into a pulse train separated by the subcavity roundtrip time $\tau_s$.  Ideally, the roundtrip time of the gain cavity is an integer $N$ times the subcavity’s, or written in free spectral ranges (FSR): 
\[  \text{FSR}_s = N \text{FSR}_g , \]
In the main experiment we are at the 10th harmonic, so $N$ = 10.  Rational harmonic modelocking is also possible with integer ratios between the roundtrip times.

The action of the subcavity can then be simulated by considering the modification of its intracavity field $X(t)$ by coupling a single input field $A_{out}$ into the subcavity
\[ X'(t) = TX(t) + RA_{out}(t)e^{i\delta\phi} ,\]
where $R$ and $T$ are the reflection and transmission coefficients in terms of amplitude (not the power coefficients) of the intracavity beamsplitter.  $X(t)$ is the field already present in the cavity at the point just below the beamsplitter in Fig. \ref{figLaser}, $A_{out}$ is the field coupled from the NALM, and $\delta \phi$ is a phase difference. 

For $X(t)$, $A_{out}(t)$, and $A_{in}(t)$ we give each array a temporal length of the gain cavity roundtrip time.  One approach to the calculation is to divide these into $N$ subsections labeled ${(m)}$ with lengths of the roundtrip time in the subcavity, $X^{(m)}$, $A_{out}^{(m)}$, and $A_{in}^{(m)}$.  The coupling of $A_{out}$ into the subcavity can be calculated from the previous subcavity field as
\[ X^{(m)} = \left( X^{(m-1)} T + R A_{out}^{(m)} e^{im\delta\xi} \right) T \]
In this model individual subsections only interact via the above equation, which is applied $N$ times.  The $N$ subsections $X^{(1)}{,\ X}^{(2)}{,\ldots,X}^{(N)}$ are concatenated to get the complete array $X(t)$ generated by the subcavity from the incident $A_{out}(t)$.  Here we also assume a  phase difference $\delta \xi$, which accumulates linearly with $m$.  Any small timing errors between  pulses can also be included, as long as the pulses stay within the temporal window of the subsections. The outputs from the subcavity are $A_{in}^{(m)}\left(t\right)=-RX^{(m)}$.  These $N$ subsections are also concatenated to build the output field $A_{in} \left(t\right)$ that will be injected back into the NALM. For the free space rejection outputs, the field subsections transmitted through the beamsplitter are $A_{free}^{\left(m\right)}\left(t\right)=\ -TA_{out}^{\left(m\right)}\left(t\right)+RX^{\left(m\right)}(t)$, which are also concatenated to build the whole output field $A_{free} (t)$.  To ensure that $A_{free} (t) \approx 0$, i.e. maximum field enhancement in the subcavity, the phase in each subsection and the input field should be the same, so there is no phase error accumulation. 

For calculation of quantum-limited timing jitter (see below), we can assume there are no phase variations between individual subsections and the input field.  In this case we find it more convenient to avoid subsections, and maintain all arrays at a length of the gain cavity roundtrip time.  We can then calculate $X{(t)}$ and $A_{in}(t)$ from an input $A_{out}(t)$ recursively:
\[ X_k\left(t\right)=\left[T X_{k-1}(t-\tau_s) + RA_{out}\left(t\right) e^{i\delta\phi} \right] T. \]
The recursive calculations are like roundtrips through the subcavity, eventually filling the whole field $X_k (t)$ in the time domain after $N$ round trips, so that $X(t) = X_N(t)$.  The field injected back into the NALM can thus be written as $A_{in}\left(t\right)=-RX_N(t)$, without relying on subsections.  Similarly, the field transmitted through the beamsplitter is $A_{free}(t) = -TA_{out}(t) + RX(t)$. To minimize any free-space output, $\delta\phi$ is selected to compensate for any phase offset from nonlinear propagation in the fiber loop.  In the experiment, we observe significant output through the beamsplitter due to imperfect mode matching between the subcavity and the fiber output and what would correspond to small residual phase variations between the input field and the individual subsections.

To reduce the number of points in the calculation we typically shrink the temporal window down to 1 to 0.1~ns or less (with corresponding change to the subcavity timing as well) while keeping the nonlinearity the same. Essentially this puts the pulses much closer together than in the real system. However, this is permissible since the pulses have a width of about 100 – 300 fs.

We found numerically that the pulse train can evolve from noise.  In experiments we used a reflectivity of ${\sim30\%}$ and the pulses generally start from a Q-switching instability. If the reflectivity of the beam splitter is too low, the large loss through the beamsplitter overwhelms the gain and the laser will not start modelocking or even cw lasing.

Once started, the steady state pulse solutions are very close to the pulse solutions for a laser system without a subcavity operating at the fundamental repetition rate, so the generated pulse forms and pulse energy are to first order independent of the harmonic number. This means average output power and pump power must increase proportionally as the harmonic number is increased.

To further increase the repetition rate, the fibers need to be lengthened to increase nonlinearity for compatibility with lower pulse energies, increasing the fundamental repetition rate of the gain cavity.  For example, we have achieved 1.5~GHz repetition rate with only about 1~W of pump power with a gain cavity fundamental repetition rate of around 35~MHz. Hence from simple scaling considerations, repetition rates up to 3~GHz should at least be reachable. Higher repetition rates would require more power by using cladding pumping, or much longer fibers.

\section{\label{secJitter}Timing jitter}

Adding a subcavity to the laser ideally adds no loss, as discussed above.  Since our experiment did not reveal any measurable supermode noise, it is instructive to evaluate the influence of quantum noise on the performance of the present system. Quantum noise from amplified spontaneous emission (ASE) will be the same for a harmonically modelocked fiber laser compared to the same fiber laser operating at its fundamental repetition rate, since quantum noise is added only once per gain cavity roundtrip \cite{SMpaschottaAPB2006} and pulse energy does not increase with harmonic modelocking with a fiber laser.  While a conventional end mirror would reflect noise back to the amplifier, the interference condition at the beamsplitter continuously filters the laser, directing a large fraction of ASE out of the beamsplitter rejection port, and turning the subcavity into a low noise (ideally noise free) pulse reservoir, greatly reducing the overall timing jitter.

\begin{figure}[b]
\includegraphics[width=89mm]{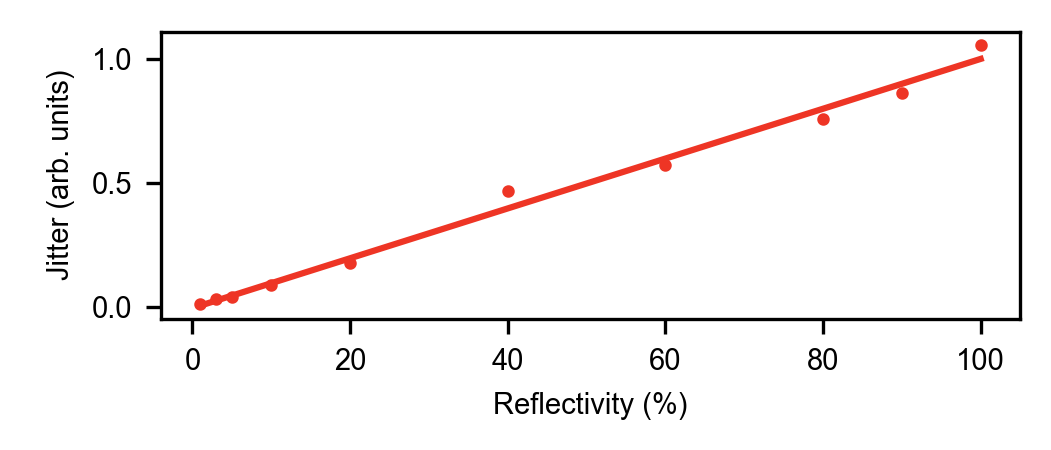}
\caption{\label{figSMjitter}
Numerical estimate of fundamental timing jitter reduction of a harmonically modelocked fiber laser as a function of intracavity beamsplitter reflectivity compared to a laser without a subcavity. 
}\end{figure}

Based on our model above, we have evaluated the expected fundamental timing jitter, adding the quantum noise term $P_{ASE}$ once per round trip time of the gain cavity.  Specifically, the noise term was added before reinjection of the pulse train into the NALM loop, using a modified term
\[ A_{in} = -R X_m + N_{ASE} . \]
$N_{ASE}$ is a noise term with variance $\frac{h\nu}{\delta t}(G-1)$ \cite{SMpaschottaAPB2004}.  The result of this simulation is shown in Fig. \ref{figSMjitter}. The fundamental timing jitter is directly proportional to the reflectivity $R$ of the intracavity beamsplitter, which also governs the power stored in the subcavity, which is proportional to $1/R$, assuming no additional losses inside the subcavity.  Per round trip through the gain cavity, a fraction of the power stored in the cavity leaks out of the subcavity and is returned back to the main cavity.  For 100\% reflectivity, there is no power enhancement and the fundamental timing jitter is the same as for a laser without a subcavity.

Clearly it is advantageous to operate the system with as low a reflectivity as possible to obtain the lowest noise performance; with a beam splitter reflectivity of 10\% a jitter reduction by a factor of 10 should be possible, which corresponds to a phase noise reduction of 20 dB compared to a standard modelocked laser. Indeed, we have demonstrated that harmonic modelocking can be stable with a beamsplitter reflectivity of only 10\%, corresponding to a cavity $Q$ of $1.2\times10^7$. 

A hard limit for the maximum possible Q for the reference cavity is given by the maximum gain in the erbium fiber laser, which can reach 40 dB, corresponding to a 1\% beam splitter reflectivity. However, operating with such a small coupling between cavities is difficult because it is almost like the laser is missing an end mirror.  The build-up time of power in the cavity becomes very long, and there is higher sensitivity to phase mismatch between the cavities.

\section{\label{secLaser}Laser description}

The main GHz laser is illustrated in Fig. 1a of the main text.  The system is fully polarization maintaining.  The gain fiber is a 44 cm length of core-pumped, small core, highly doped Er fiber (Liekki Er80-4/125-HD-PM).  It is pumped by two 1 W laser diodes using polarization combining.  One of the diodes can be modulated at rates over 100 kHz.  The main output is from a 99/1 coupler in the NALM loop, with about 1 mW of power.  The NALM rejection port outputs about 30 mW, and the free space rejection port outputs about 50 mW.

The f-2f interferometer used the oscillator tap output, and amplified it in three core-pumped amplifiers (exail IXF-EDF-FGL-PM fibers) with a total of five 1 W pump diodes to bring the beam close to 1 W average power, compressible to about 63 fs.  The supercontinuum was generated in 1.6 m of highly nonlinear fiber (Thorlabs PMHN5) \cite{SMleskoOSAC2020}.  The f-2f beat note had about 35 dB signal to noise ratio at 100 kHz resolution bandwidth.  Further fiber length optimization should be possible, since the amplifier pumping was reduced significantly when optimizing the f-2f signal.  A cladding pumped amplifier may be more practical.  f-2f interferometry has been demonstrated with high average power at 16.4 GHz \cite{SMdelhayeNPh2016}.  Another option is to use chip devices for supercontinuum generation that require less pulse energy \cite{SMgrayA2025} to reduce the average power needed.

\begin{figure}[t]
\includegraphics[width=89mm]{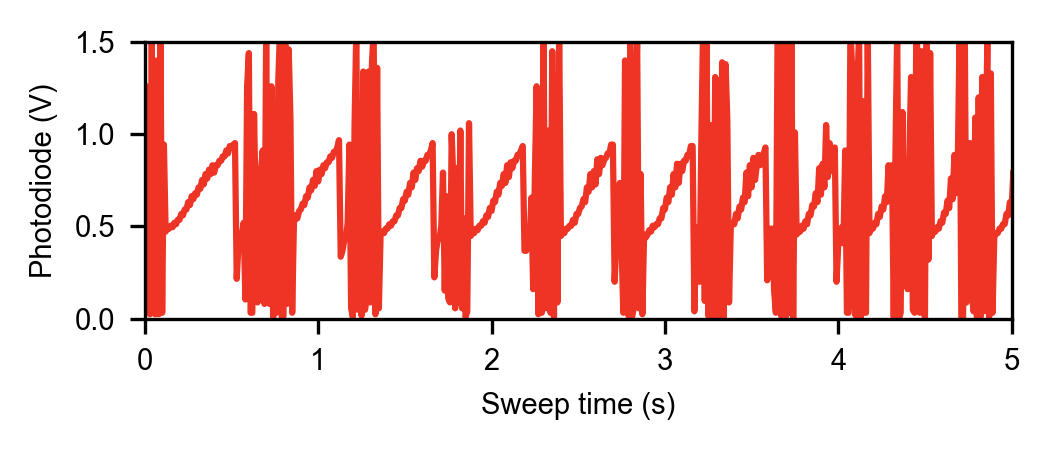}
\caption{\label{figSMsweep}
Laser output as length of gain cavity is swept in one direction by the fiber stretcher.  Resonances are separated by a wavelength.  The laser operates over a portion of a wavelength (smoother sections), and is unstable in the other regions.
}\end{figure}

The gain cavity had a free spectral range of around 104 MHz, and we usually operated at the 10th harmonic.  Selecting lower harmonics only requires lengthening the subcavity by moving the end mirror.  Higher harmonics are likely power limited, as the pumps run at nearly full power at the 10th harmonic.  Since the laser operated quite easily, the free space optics were not optimized for low loss beyond adding a lens in the subcavity, so the pumping requirements are likely lower than what we used here.

\begin{figure}[t]
\includegraphics[width=89mm]{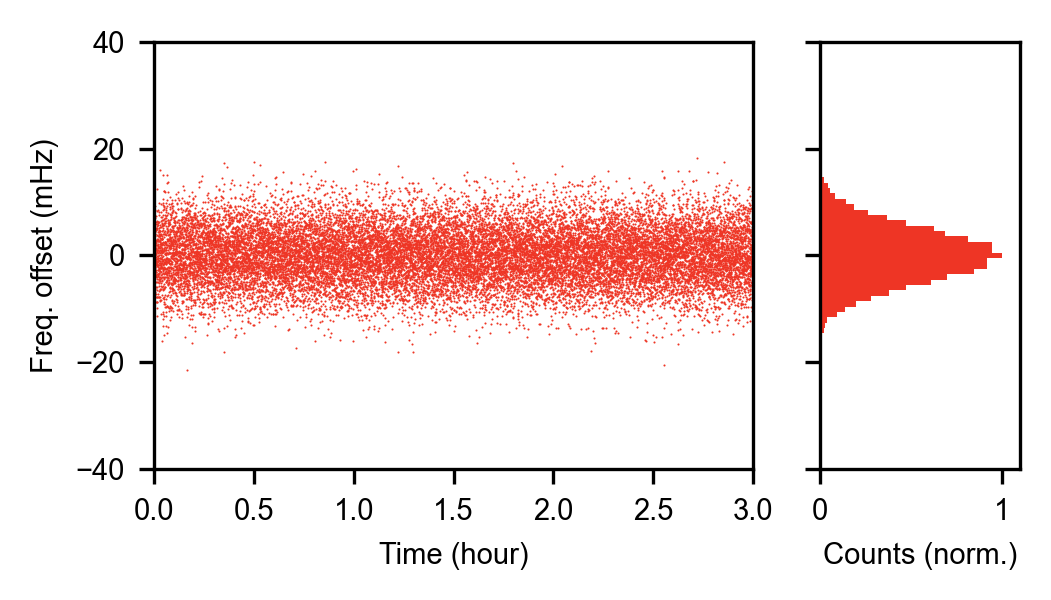}
\caption{\label{figSMmixedBeat}
Stability over 3 hours of the out of loop comparison between comb lines from the GHz and reference combs, with histogram shown on the right.  The frequency stays within about 15 mHz of the average value.
}\end{figure}

For setting cavity lengths, first we coarsely move the subcavity mirror, observing noisy output bursts when passing a harmonic resonance.   The burst is actually a set of resonances spaced by a wavelength.  When scanning the fiber stretcher near resonance, the laser will go in and out of modelock.  A measurement of the free space rejection output while ramping the fiber stretcher is shown in Fig. \ref{figSMsweep}.  The smoother regions correspond to pulsing, which includes both clean and noisy harmonic modelocking, while the spiking regions are like Q-switching.

The CEO and optical beat notes are very tunable, and can move across the entire 1 GHz span, although not continuously, as the laser will go in and out of modelock while scanning the fiber length.  To achieve specific frequencies, we coarsely adjust the subcavity length for the CW beat note, and the pump current for the CEO frequency before locking.

Simply changing the fiber length is sufficient to modelock. Modelocking will change the fiber temperature, so active adjustment is needed in the first few seconds, at about 10 Hz bandwidth.  After warm up, the laser may operate passively for many minutes, but active feedback is generally required.

\section{\label{secLoop}Out of loop stability}

We show an additional plot in Fig. \ref{figSMmixedBeat} related to the out of loop comb comparison.  It shows the stability of the mixed beat note over 3 hours.  The frequency difference between the GHz and reference comb lines stays within about 15 mHz of the average value.  Each point is the average of the mixed frequency in 0.5 s.

\section*{}\

\end{document}